\documentclass[twocolumn,showpacs,preprintnumbers,amsmath,amssymb,prl,superscriptaddress,nofootinbib]{revtex4}

\usepackage{graphicx}
\usepackage{dcolumn}
\usepackage{bm}

\begin{document}
\title{Phase behavior and far-from-equilibrium
gelation of charged attractive colloids}
\author{P. Charbonneau}
\affiliation{Chemistry and Chemical Biology, Harvard
University, 12 Oxford Street, Cambridge, Massachusetts 02138,
USA~\footnote{Current address: FOM Institute for Atomic and
Molecular Physics, Kruislaan 407, 1098 SJ Amsterdam, The
Netherlands.}}
\author{D. R. Reichman}
\affiliation{Department of Chemistry, Columbia University, 3000
Broadway, New York, New York 10027, USA}

\pacs{82.70.Dd, 64.60.-i, 61.20.Lc, 61.46.Bc}
\date{\today}
\begin{abstract}
In this Rapid Communication we demonstrate the applicability of
an augmented Gibbs ensemble Monte Carlo approach for the phase
behavior determination of model colloidal systems with
short-ranged depletion attraction and long-ranged repulsion.
This technique allows for a quantitative determination of the
phase boundaries and ground states in such systems. We
demonstrate that gelation may occur in systems of this type as
the result of arrested microphase separation, even when the
equilibrium state of the system is characterized by compact
microphase structures.
\end{abstract}
\maketitle

Understanding the routes to colloidal gel formation is a
problem in the forefront of soft-condensed matter physics.  It
has been demonstrated via theory~\cite{cates:2004},
experiment~\cite{verhaegh:1997,manley:2005}, and
simulation~\cite{foffi:2005,foffi:2005b} that a weak, porous
solid may be formed when a suspension of colloidal particles
with short-ranged attractions is quenched below its critical
point.  Spinodal decomposition, expected to lead to complete
phase separation into colloid-rich and colloid-poor regions,
may arrest or become anomalously slow when bonding between
colloidal particles is sufficiently strong. The resulting
structure may support weak shear stresses, and is called a
colloidal gel. While equilibrium routes to physical colloidal
gelation do
exist~\cite{zaccarelli:2005,delgado:2005,shah:2003}, the
nonequilibrium route described above is perhaps the most
ubiquitous.

When excess charge resides on the colloidal particles, the
situation is more complex~\cite{segre:2001}. In addition to
short-ranged depletion interactions, the long-ranged repulsive
portion of the potential may lead to the formation of various
microphase structures, such as clusters, cylinders, sheets, and
spirals~\cite{mossa:2004,sciortino:2005,sear:1999b,imperio:2004,deCandia:2006}.
As a result, the nature of physical gelation in these systems
is potentially more subtle than in systems with attractive
interactions alone. Indeed, several novel routes to gelation in
these systems have been discussed. One proposal is that
compact, thermodynamically stable clusters may form the
building blocks of a glassy state~\cite{sciortino:2004}.  At
high enough volume fraction clusters may aggregate or become
caged, as in the case for individual colloidal particles near
the colloidal glass transition~\cite{wu:2004}. At low volume
fractions, such a glassy state could be stabilized by the
effective long-ranged repulsion between individual
clusters~\cite{sciortino:2004}. More recently, it has been
proposed that disordered or partially ordered anisotropic
domains may be responsible for physical gelation in charged
colloidal
systems~\cite{sciortino:2005,deCandia:2006,tarzia:2006}.

A major difficulty in testing the validity of any proposal
resides in the accurate calculation of the phase diagram in
systems with long-ranged repulsive and short-ranged attractive
interactions. When the range of the attractive interaction is
comparable to the size of the particles, the phase behavior may
be characterized easily with standard simulation
techniques~\cite{sear:1999b,imperio:2004,deCandia:2006}.  On
the other hand, when the attractive portion of the potential
varies over distances that are a fraction of the particle size,
sampling issues become severe, and direct molecular dynamics
(MD) and Monte Carlo (MC) techniques are not feasible.
Unfortunately it is in the short-ranged attractive limit that
the gel phase is usually
stabilized~\cite{foffi:2005,foffi:2005b}. We have recently
demonstrated~\cite{charbonneau:2006,liu:2005} that a judicious
implementation of the Gibbs ensemble Monte Carlo
(GEMC)~\cite{panagiotopoulos:1992} technique allows for phase
behavior computation in systems with very short-ranged
attractive interactions. In this Rapid Communication we adapt
the GEMC technique for the study of systems characterized by
short-ranged attraction and long-ranged repulsion. The
successful application of this approach provides crucial
information on possible dynamical routes to gelation in systems
with competing interactions, which are examined via MD
simulation. Indeed, our conclusions and interpretations of the
gelation mechanism differ from those previously drawn from
studies of analogous systems.

\label{sect2}
\begin{figure}
\center{\includegraphics[width=0.9\columnwidth]{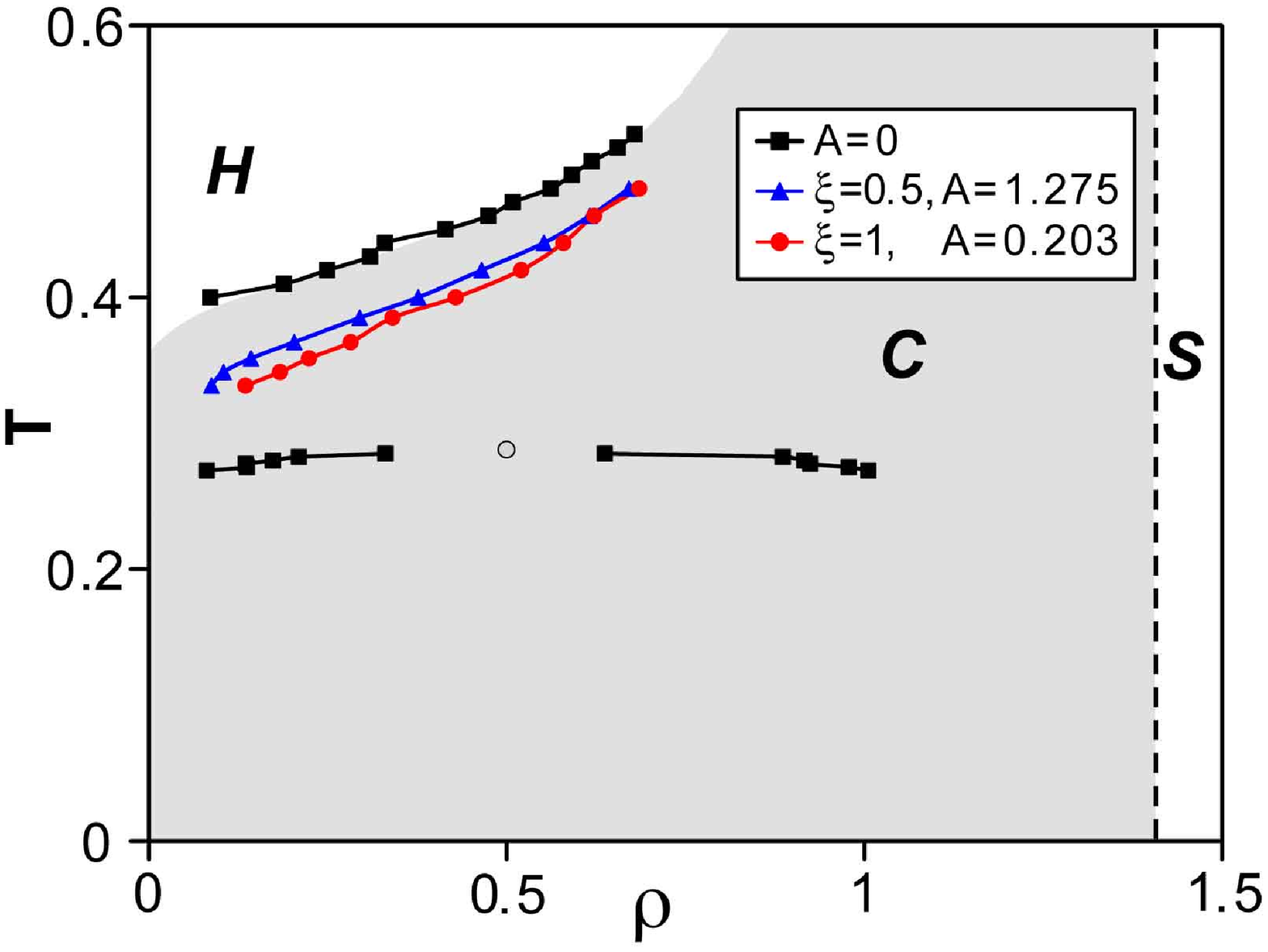}\\
\hspace{-0.1in}\includegraphics[width=0.9\columnwidth]{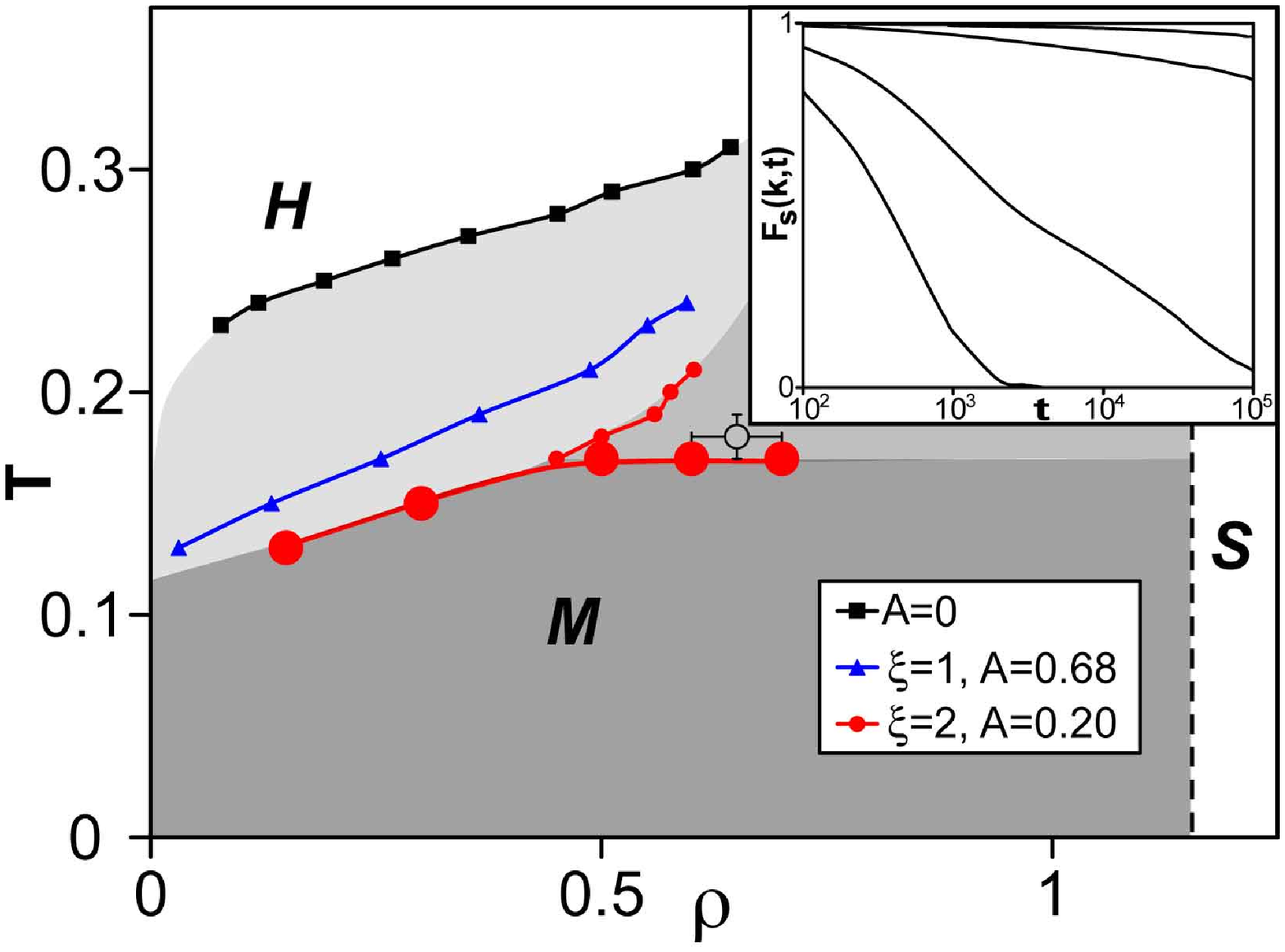}}
\caption[Two- and three-dimensional phase diagrams for
short-ranged attraction with different repulsion ranges]
{(Color online) Phase diagrams for short-ranged attraction in
(a) three ($n$=$50$) and (b) two ($n$=$100$) dimensions with
increasing repulsion set by $A$ and $\xi$. The dashed line is
the crystal close-packed density. Critical points (open
circles) for purely attractive systems ($A$=$0$) are obtained
as described in the text. Small symbols indicate gas-solid and
gas-liquid coexistence and large symbols the microphase
boundary for $\xi$=$2$ and $A$=$0.2$. Uncertainty is smaller
than the symbols, unless otherwise indicated. The phases are
labeled M (microphase), H (homogeneous fluid), C (gas-crystal
coexistence), and S (solid). Lines and shaded areas are guides
for the eye. Inset: $F_s(k,t)=\langle \exp\left\{{\mathbf
k}\cdot\left[{\mathbf r}_i(t)-{\mathbf
r}_i(0)\right]\right\}\rangle$ from MD of the microphase
forming system at the low-$k$ peak of the structure factor
$k\sigma$$\approx$$0.9$ for $\rho$=$0.6$ and $T$=$0.22$,
$0.17$, $0.12$, and $0.08$, from left to right. For the higher
two temperatures, trajectories are run for many times the
relaxation time. For the lower two temperatures the system
shows aging.} \label{fig:2dph}
\end{figure}

In this work we study potentials of the
form\footnote{Temperature $T$ is in units of well depth
$\epsilon$, distance in particle diameter $\sigma$, and time
$t$ in $\sqrt{\epsilon/m \sigma^2}$ with particle mass
$m$.}~\cite{sciortino:2004}
\begin{equation}
U(r)=4\epsilon
\left[\left(\frac{\sigma}{r}\right)^{2n}-\left(\frac{\sigma}{r}\right)^n\right]
+A\frac{e^{-r/\xi}}{r/\xi}\textrm{,}
\end{equation}
where large $n$ values correspond to short-ranged attractions.
We study three-dimensional systems ($n$=$50$) with three sets
of repulsive Yukawa parameters
($A$=$0$~\cite{charbonneau:2006}; $A$=$1.275$, $\xi$=$0.5$;
$A$=$0.203$, $\xi$=$1$) and two-dimensional systems ($n$=$100$)
with parameters ($A$=$0$; $A$=$0.68$, $\xi$=$1$; $A$=$0.2$,
$\xi$=$2$). For the last parameter set, the system is beyond
the Lifshiftz point and thus microphase separates at low
temperatures. A preliminary dynamical study of a related
three-dimensional system has been performed by Sciortino
\emph{et al.}~\cite{sciortino:2004}. For visualization
purposes, we focus on a two-dimensional system. As discussed
below we expect all qualitative conclusions to be unmodified by
this choice. Selected results will also be provided for the
three-dimensional case.

For the vapor-solid equilibrium, we use the GEMC methodology
described in Ref.\footnote{Local and non-local moves are
equally frequent at low densities, high temperatures, and for
gas-liquid equilibria. Otherwise, non-local moves comprise up
to $90\%$ of the total. Local moves are particle displacements
($94\%$), cluster displacements ($0.5\%$), rotations ($0.5\%$),
and cleaving~\cite{whitelam:2005} (up to $1\%$), as well as
symmetric ($3\%$) and asymmetric ($1$ to $2\%$) volume
exchanges. Non-local moves are particle exchanges ($80\%$) and
one or two-box ($10\%$ each) aggregation volume
bias~\cite{chen:2000}. }~\cite{charbonneau:2006}. Initial boxes
contain 256 particles in the gas phase and 500 particles in the
solid slab. A minimum of $10^6$ MC cycles are performed for
equilibration and production. To characterize the phase
behavior in the regime where microphase separation occurs, two
identical boxes with a minimum of 256 particles on a lattice
are used for initial configurations. At these lower
temperatures, the GEMC method serves as an efficient phase
space sampling algorithm akin in spirit to the grand canonical
MC and parallel-tempering MC used in previous studies of
systems with competing attractive and repulsive
interactions~\cite{imperio:2004}. It should be noted, however,
that the approach implemented here is more efficient at
detecting complex phase boundaries. In particular, upon
crossing from a gas-solid coexistence region to a microphase
separated region, the parallel-tempering MC sampling efficiency
is dramatically reduced.

\begin{figure*}
\begin{tabular}{c|c}
  \includegraphics[width=0.3\columnwidth]{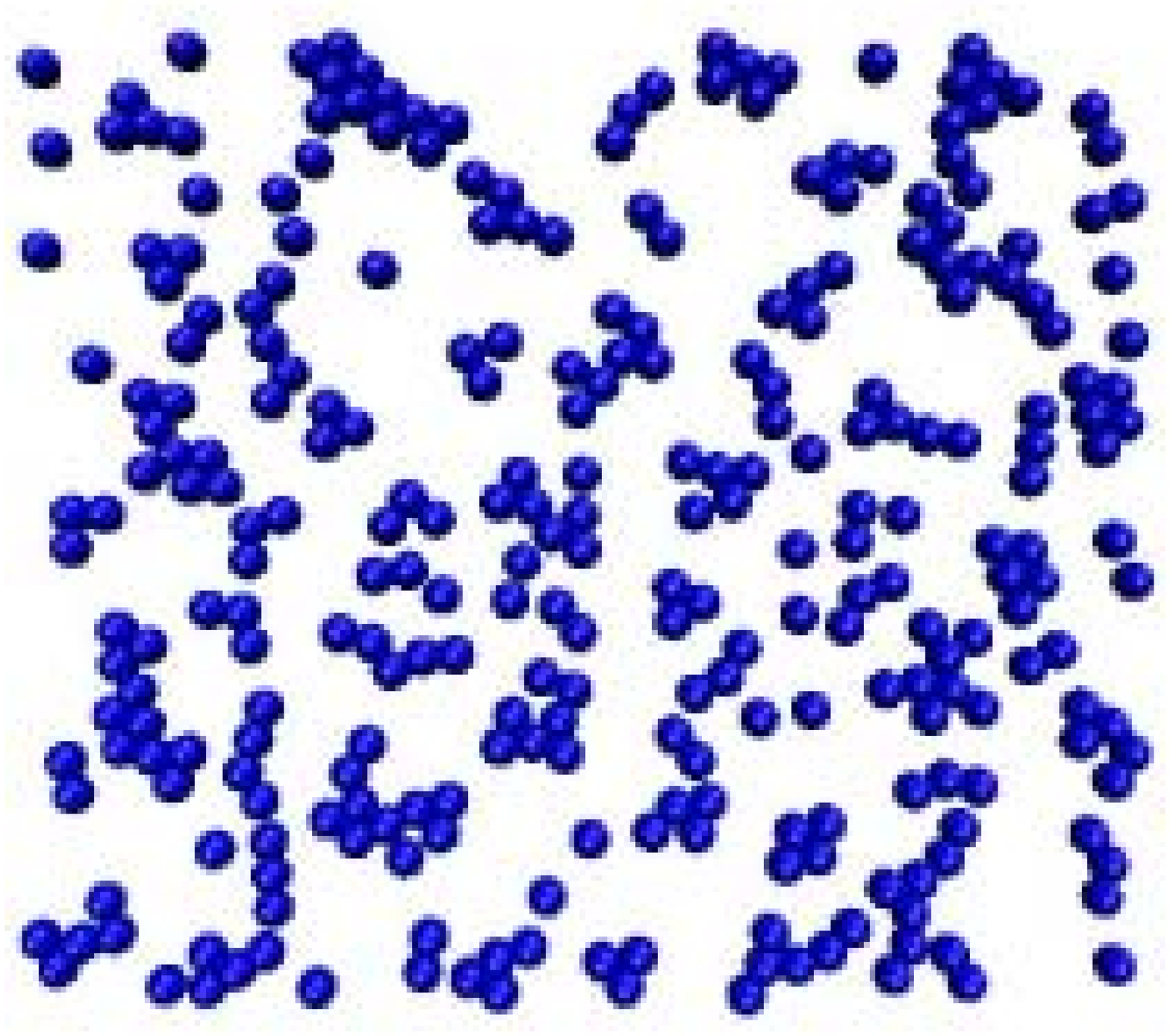}\includegraphics[width=0.3\columnwidth]{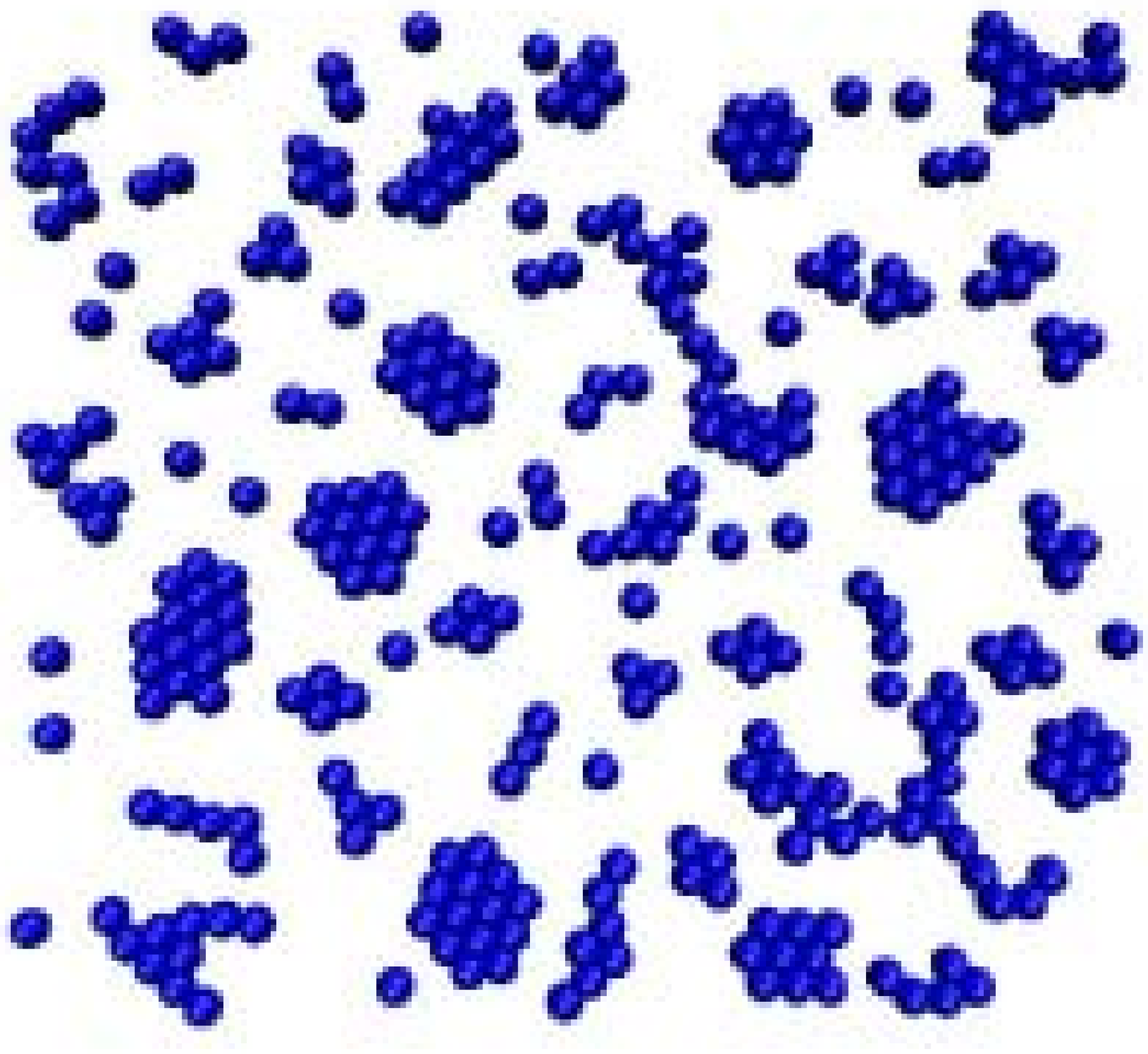}\includegraphics[width=0.3\columnwidth]{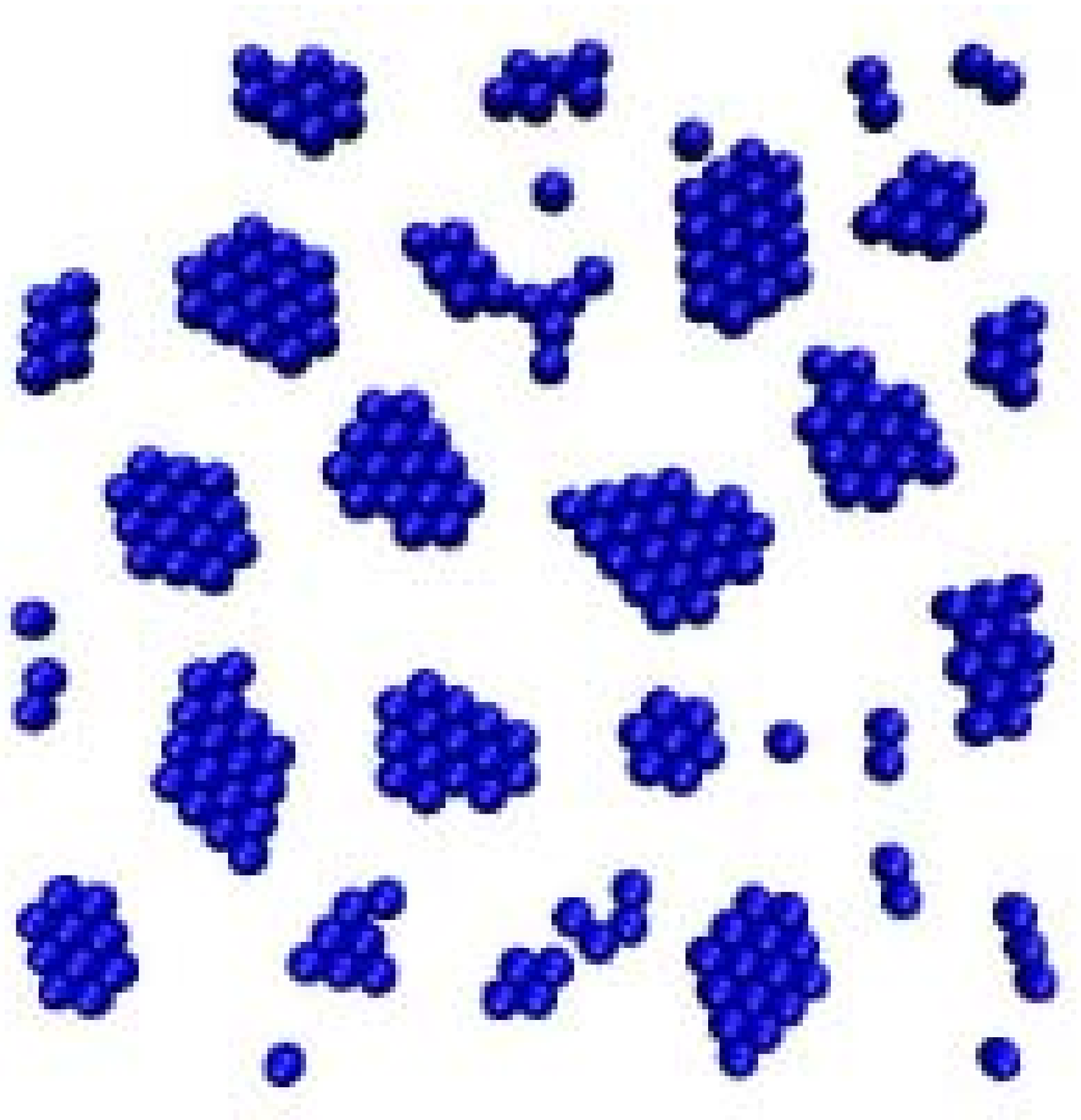} & \includegraphics[width=0.3\columnwidth]{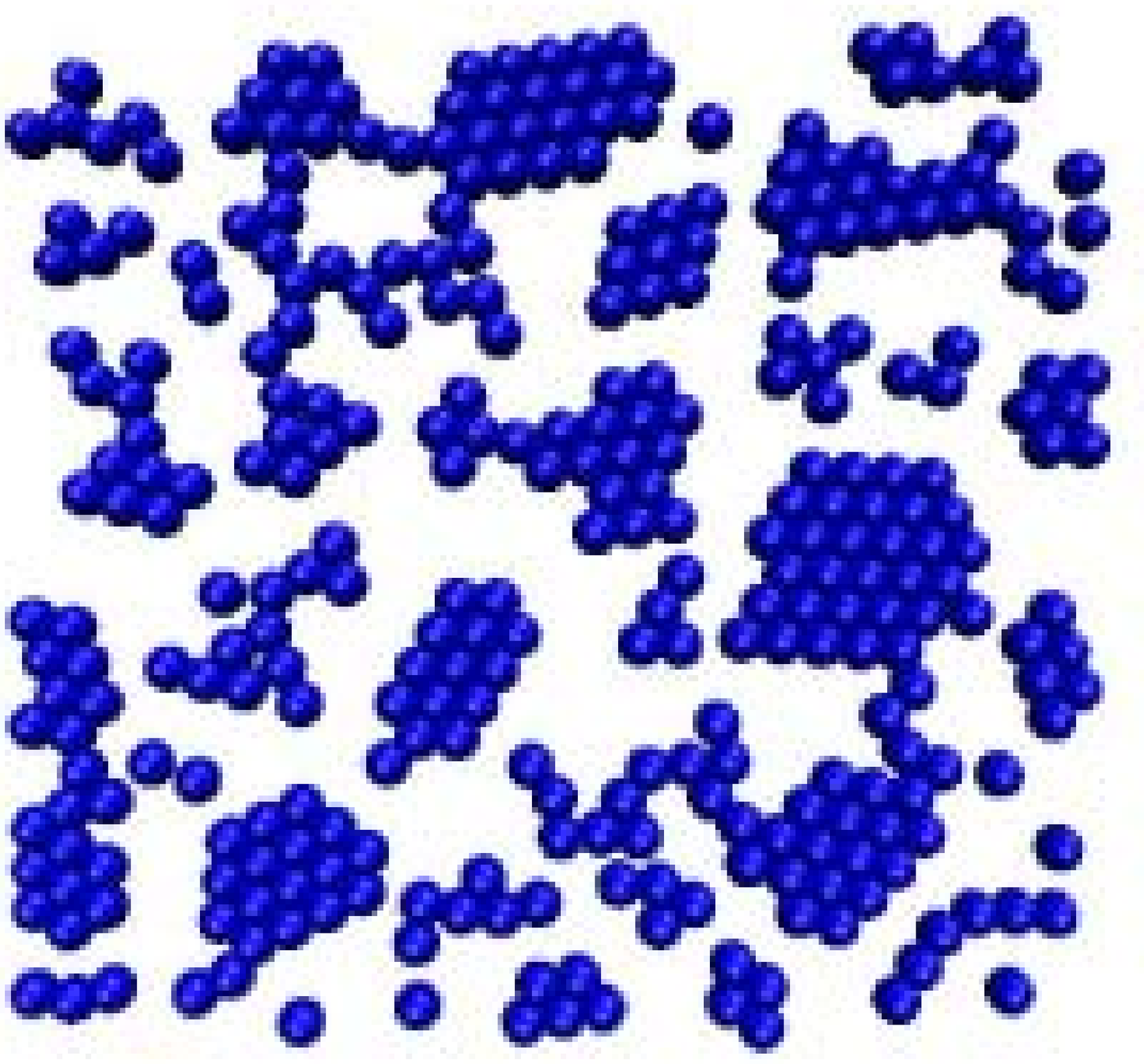}\includegraphics[width=0.3\columnwidth]{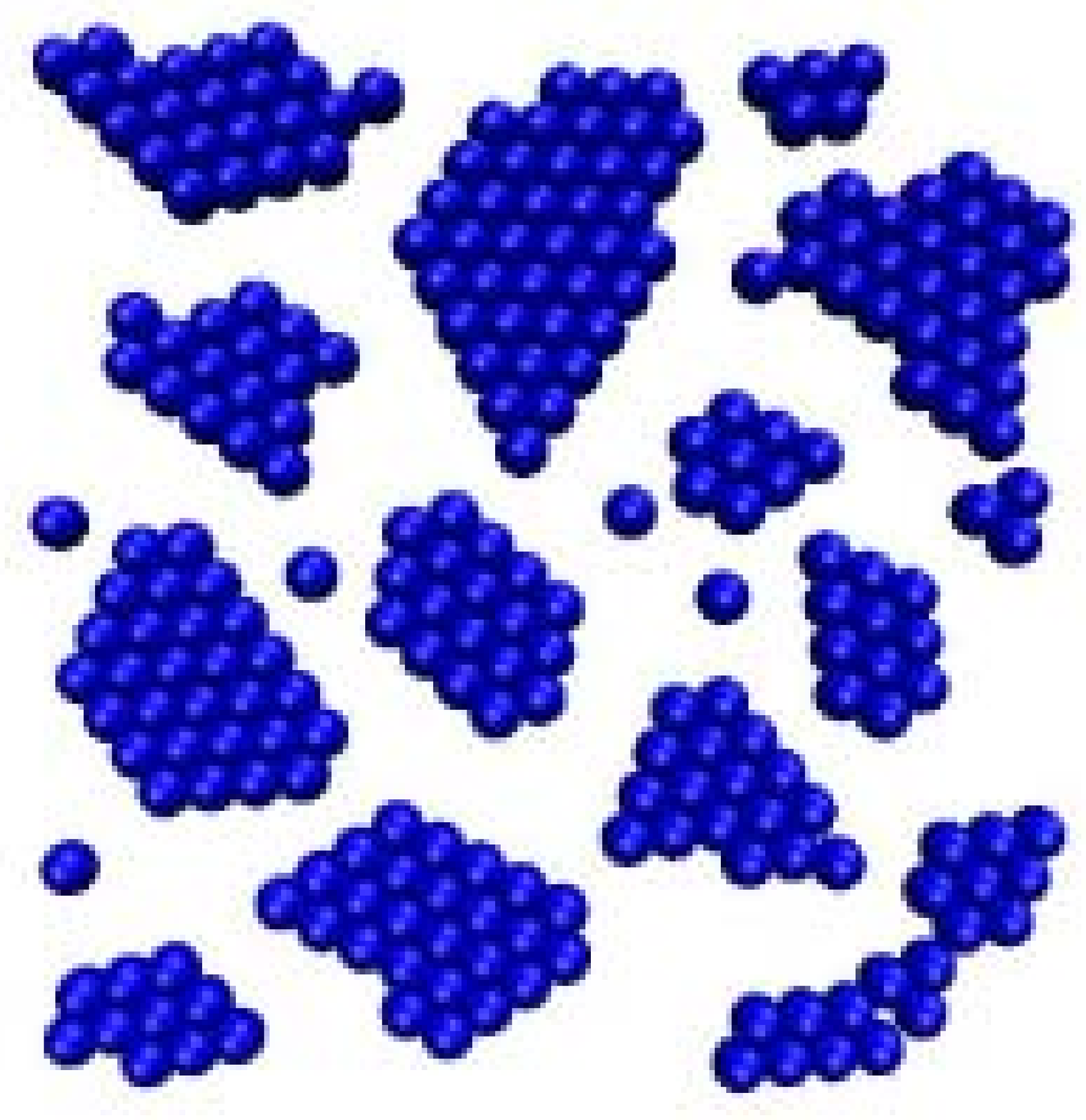}\includegraphics[width=0.3\columnwidth]{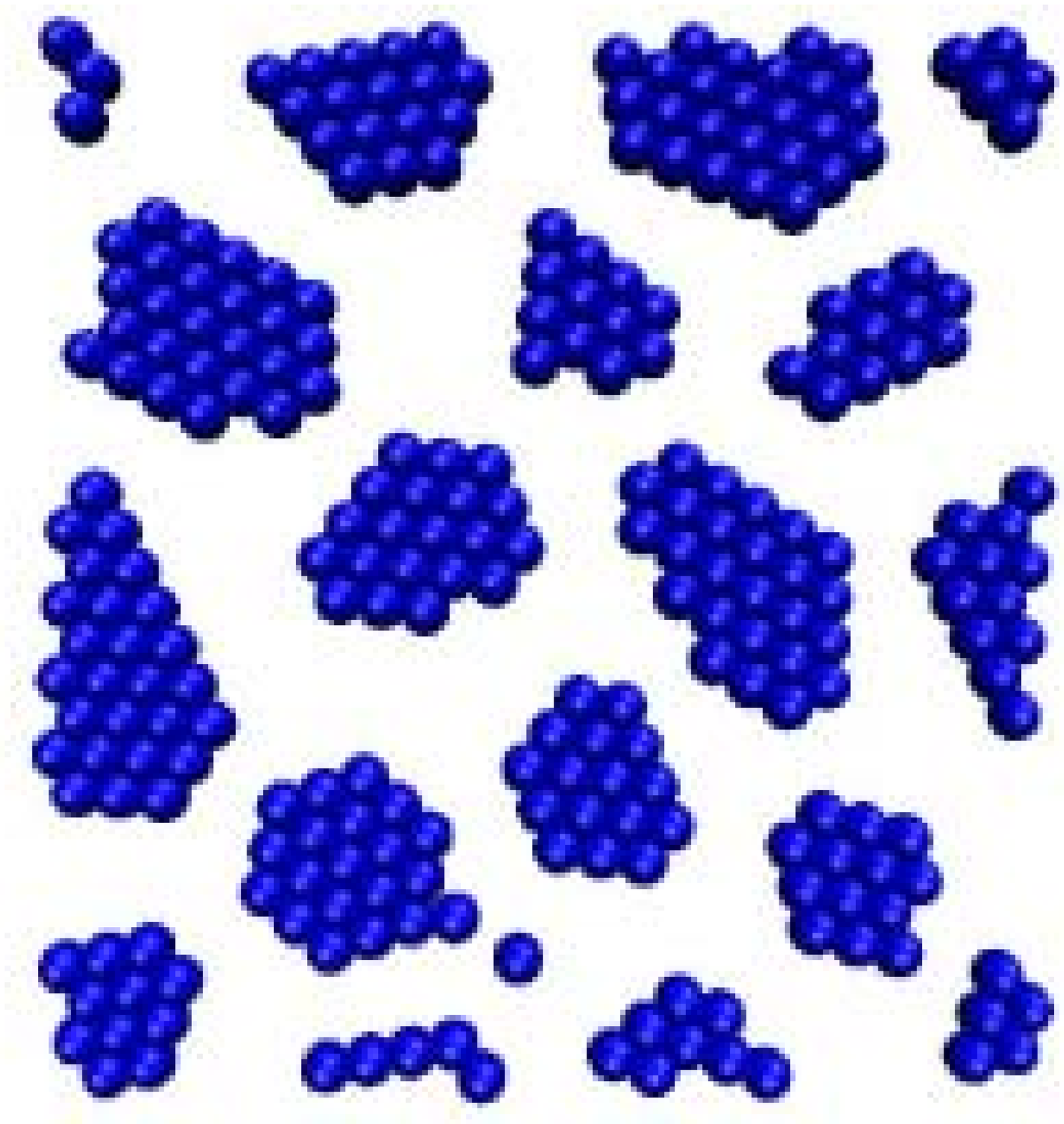} \vspace{-0.1in}\\
  (a)\hspace{.3in}T=0.16\hfill T=0.14\hfill T=0.12\hspace{.4in} & (b)\hspace{.3in}T=0.16\hfill T=0.14\hfill T=0.12\hfill\hspace{.4in}\\\hline
  \includegraphics[width=0.3\columnwidth]{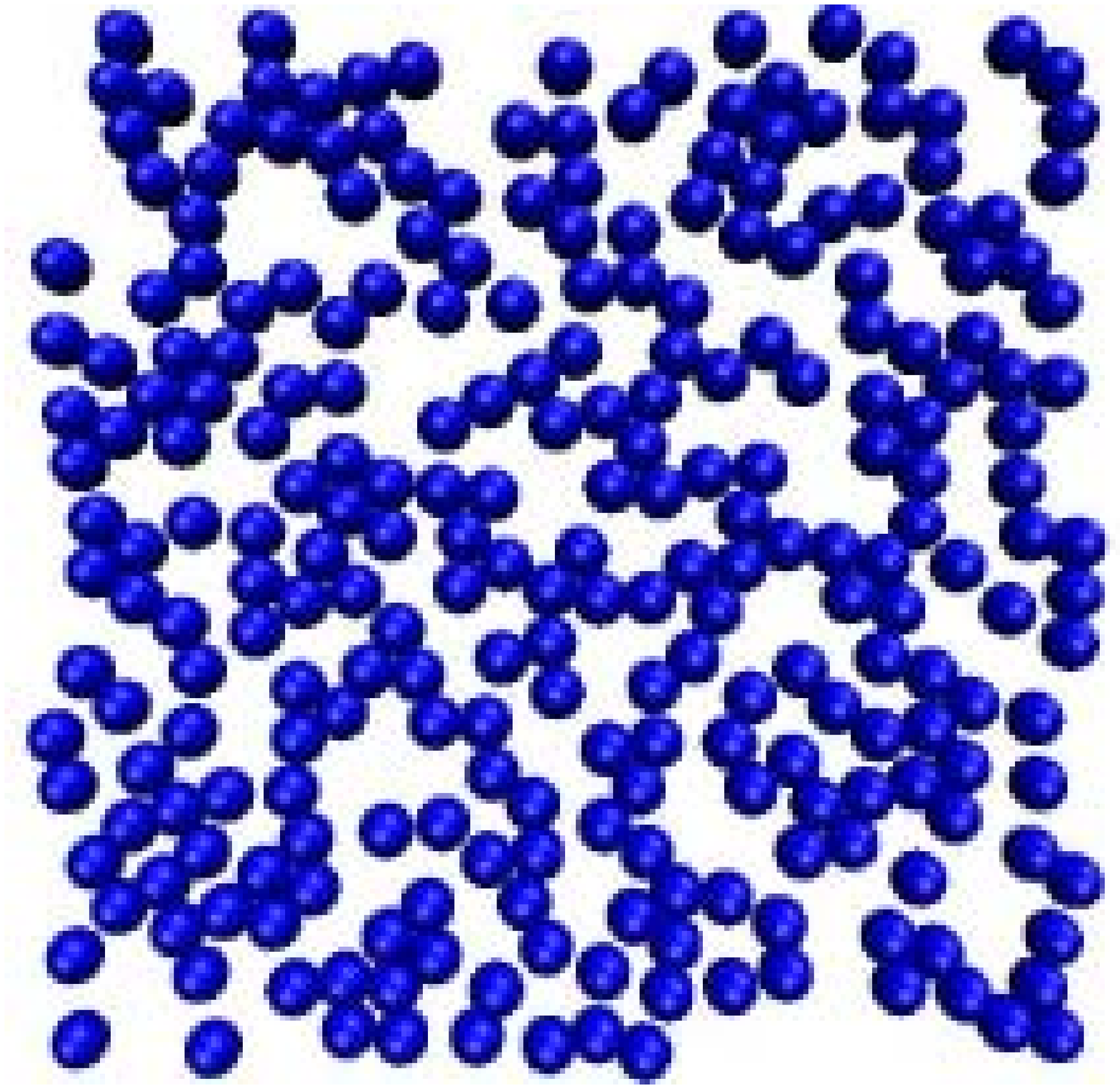}\includegraphics[width=0.3\columnwidth]{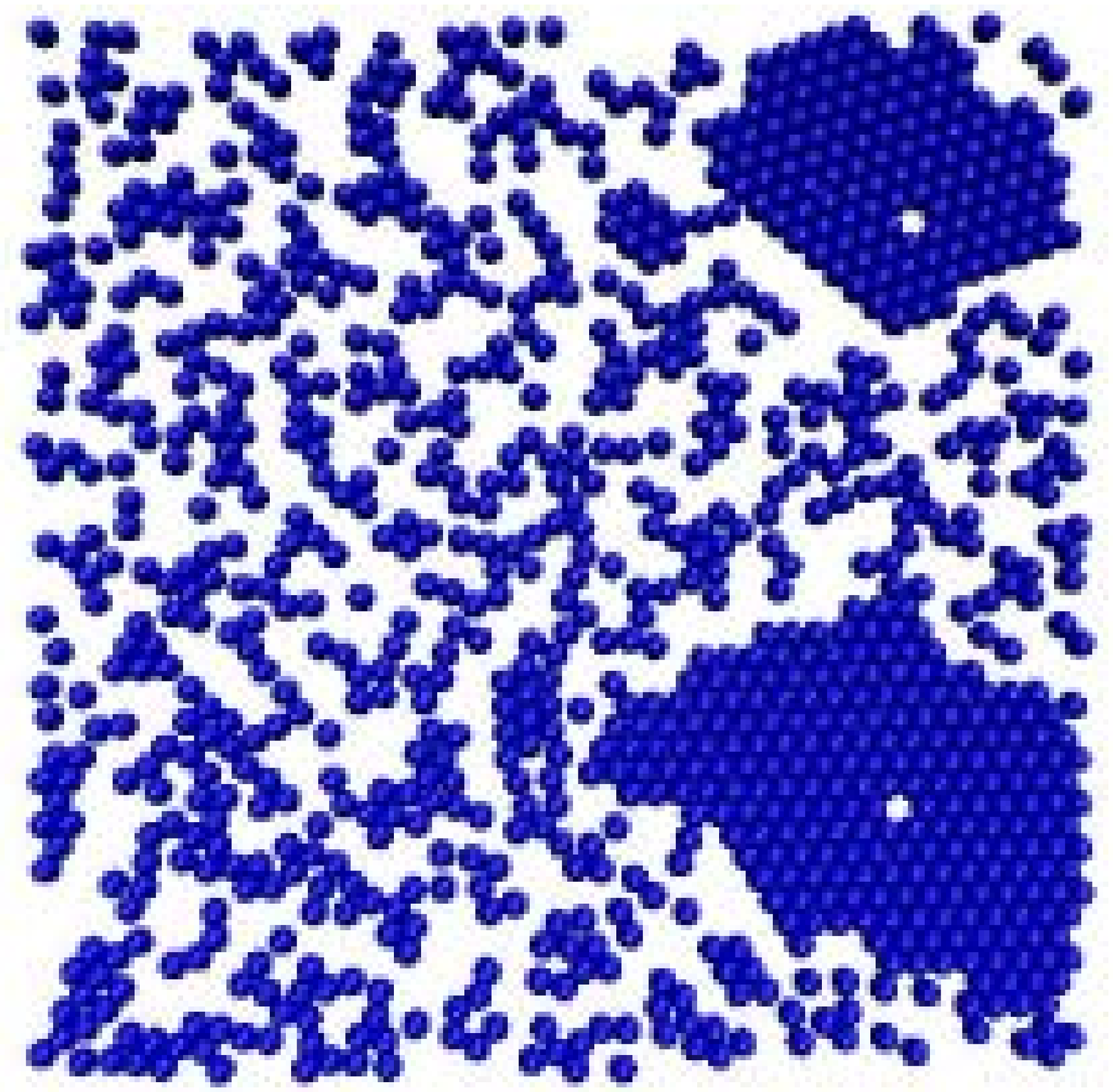}\includegraphics[width=0.3\columnwidth]{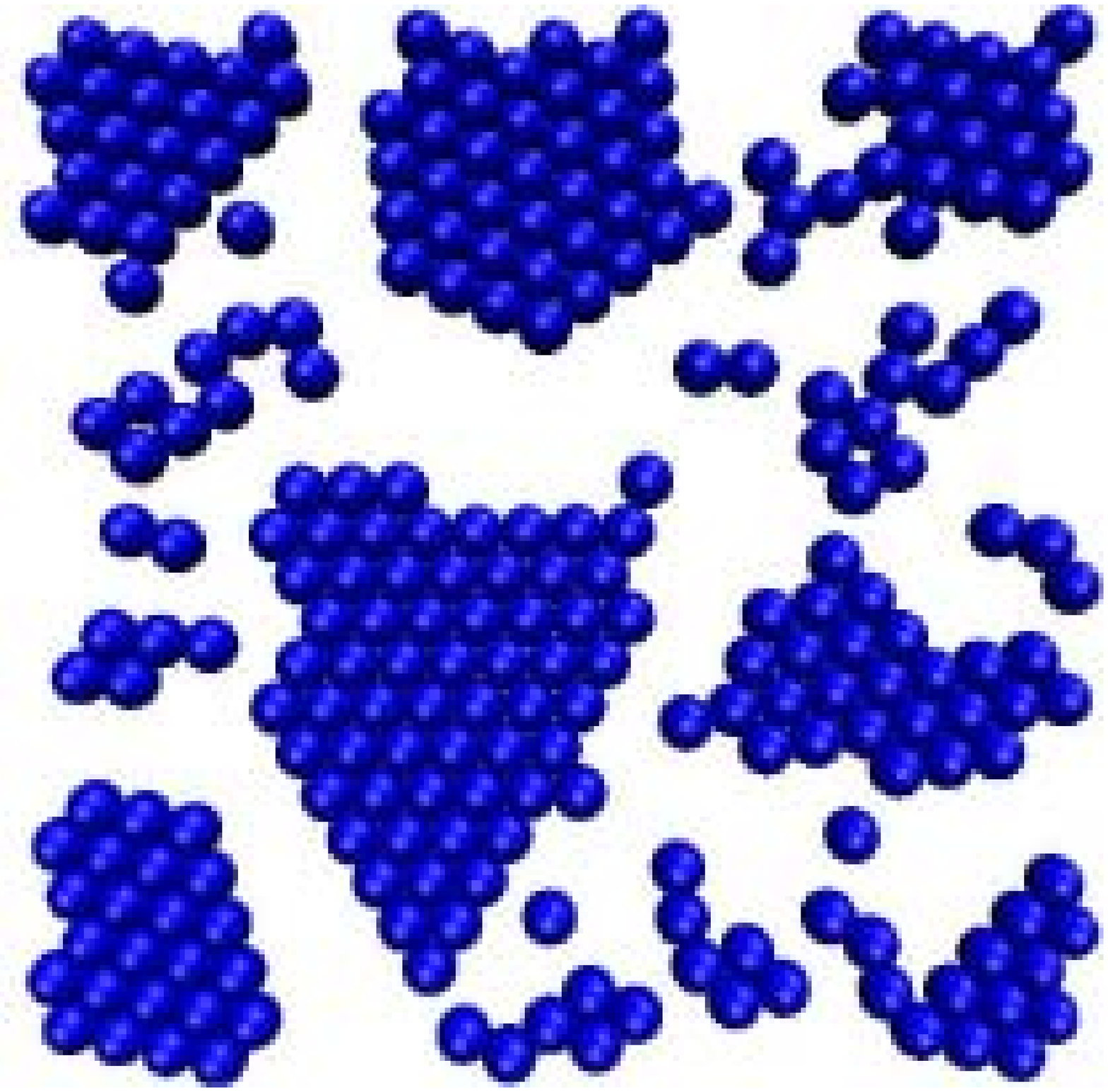} & \includegraphics[width=0.3\columnwidth]{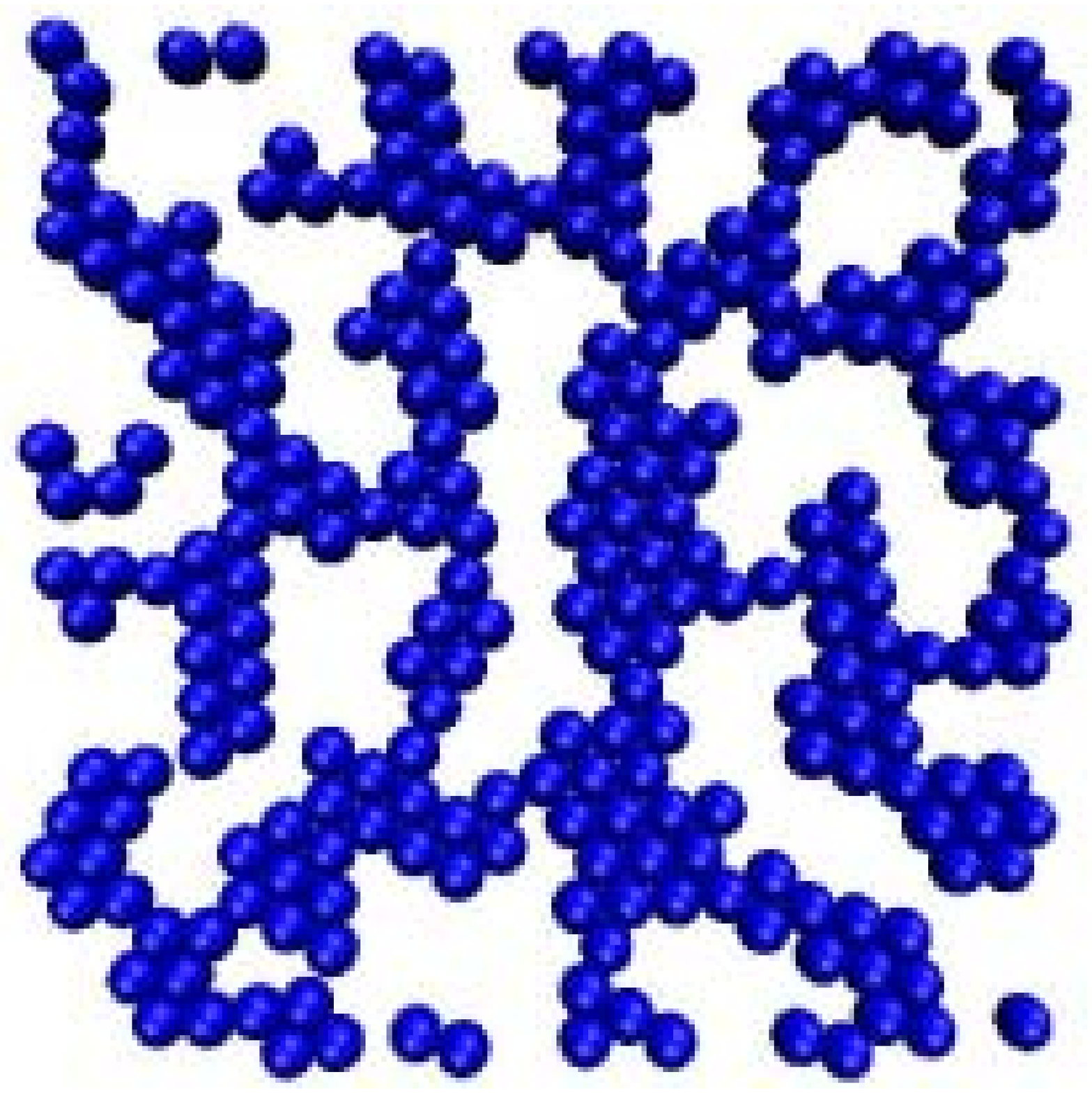}\includegraphics[width=0.3\columnwidth]{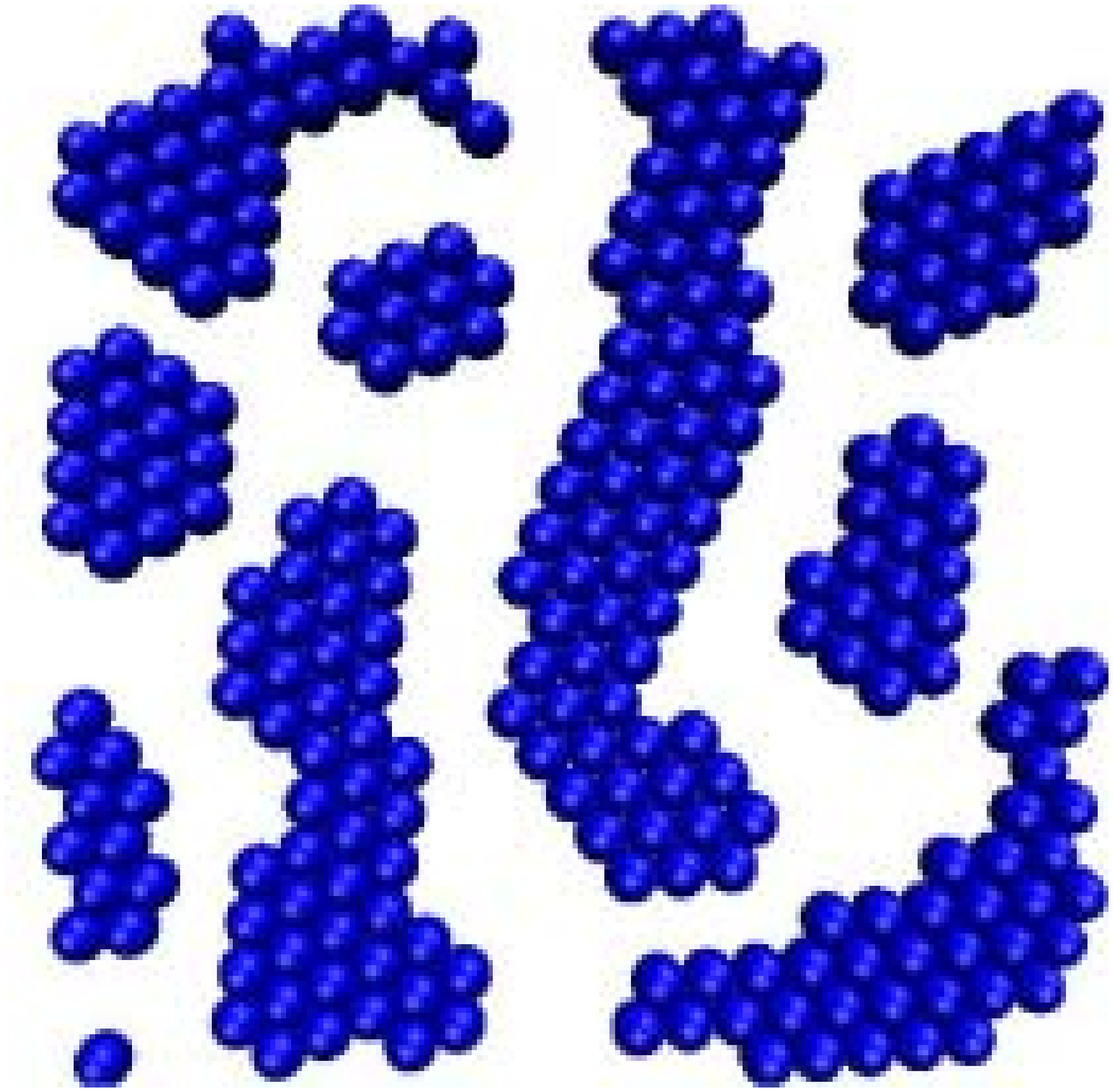}\includegraphics[width=0.3\columnwidth]{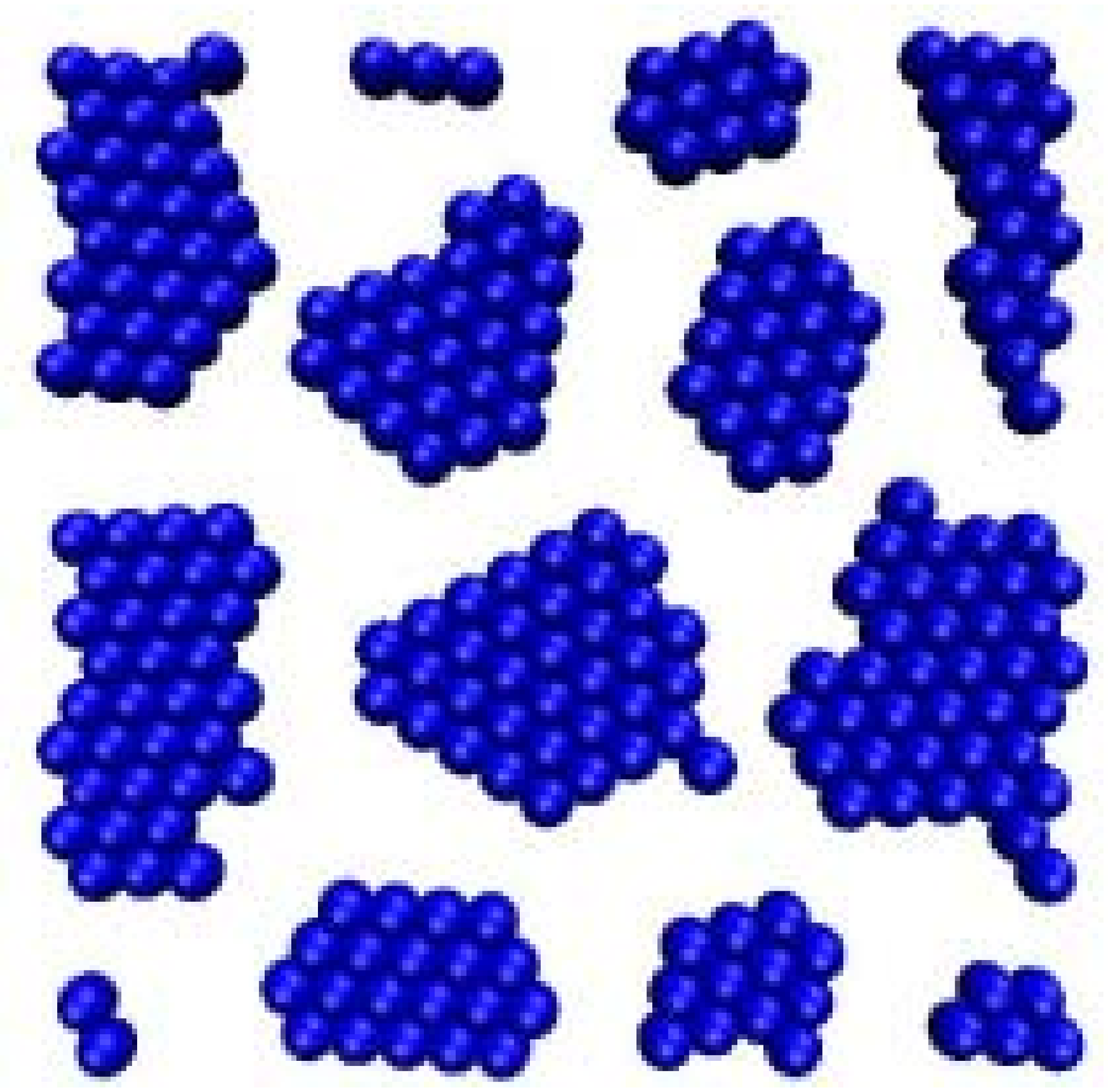} \\
  (c)\hspace{.3in}T=0.22\hfill T=0.18\hfill T=0.16\hspace{.4in} & (d)\hspace{.2in}$t=1.8\times10^4$\hfill$t=2.05\times10^6$\hfill$t=\infty$\hspace{.3in}\hfill\\
\end{tabular}
\caption[Two-dimensional GEMC thermodynamic ground state and
dynamic configurations of microphase-forming systems]{(Color
online) Configurations ($\xi$=$2$, $A$=$0.2$)
equilibrated with GEMC simulations at (a) $\rho$=$0.3$,
(b) $\rho$=$0.5$,  and (c) $\rho$=$0.6$ as well as
(d) $\rho$=$0.6$ MD evolution after quench to $T$=$0.12$ (first
two panels) and GEMC thermodynamic outcome (third
panel).}\label{fig:2dconfig}
\end{figure*}

Before discussing the impact on phase equilibrium of the
repulsive portion of the potential, we make some brief comments
on the purely attractive case in two dimensions. The phase
diagrams of short-ranged attractive two-dimensional systems
have not been investigated as thoroughly as their
three-dimensional counterparts. The existence of high-density
crystal-crystal coexistence in both two and three dimensions
supports the notion that systems characterized by identical
interactions in two and three dimensions are qualitatively
similar~\cite{frenkel:1996}. While crystal-crystal coexistence
does not occur in the temperature and density range of interest
here, this phenomena is related to the metastability of the
gas-liquid critical point, which does occur at densities and
temperatures of concern in this work. With no repulsion
($A$=$0$), we thus expect the liquid-gas coexistence to become
metastable for $n\gtrsim12$, so for $n$=$50$-$100$ the
gas-liquid binodal is metastable and buried below the gas-solid
coexistence line~\cite{hasegawa:1997}. In three dimensions it
is known that nucleation dynamics are dramatically influenced
by this buried gas-liquid
binodal~\cite{tenwolde:1997,gast:1983,hagen:1994}. Similar
effects also appear in two dimensions, although metastable
states are harder to observe, due to much faster nucleation. In
the two-dimensional attractive case ($n$=$100$, $A$=$0$), our
visual inspection of the relaxation dynamics accompanying
apparent spinodal decomposition gives $T_c$=$0.18\pm0.01$. This
can be favorably checked by combining the arguments developed
by Noro and Frenkel regarding corresponding states for
short-ranged potentials~\cite{noro:2000} with the location of
the critical point coordinates extracted from the results of
Seaton and Glandt for the Baxter limit of adhesive
disks~\cite{seaton:1986}, which further give an estimate of the
critical density $\rho_c$=$0.6\pm0.1$. For the same potential
form ($n$=$100$, $A$=$0$), but in three dimensions,
$T_c$=$0.235\pm0.005$~\cite{sciortino:2004} and
$\rho_c$$\sim$$0.5$~\cite{charbonneau:2006,miller:2004}, while
the three-dimensional system studied here ($n$=$50$, $A$=$0$)
has $T_c$=$0.288\pm0.005$~\cite{charbonneau:2006}. The
metastable critical points are indicated in
Fig.~\ref{fig:2dph}.

We now turn to a discussion of the impact of repulsion on phase
behavior. In three-dimensional systems with attractive ranges
similar to that of of the $n$=$6$ (Lennard-Jones) potential,
the addition of small amplitude long-ranged repulsion is known
to slightly depress the gas-liquid spinodal and binodal
lines~\cite{pini:2000}. Repulsion of greater amplitude or
longer range transforms the depressed gas-liquid coexistence
line into a microphase boundary where low- and high-density
regions locally coexist, as in Fig.~\ref{fig:2dph}(b). Besides,
repulsion destabilizes the crystal phase and thereby extends
the stability of the homogeneous fluid phase to lower
temperatures, as seen in Fig.~\ref{fig:2dph}. Where microphase
separation occurs, this depresses the low-density microphase
boundary. In this case the high-density phase is crystalline,
since the microphase boundary is below the gas-solid
coexistence line. GEMC temperature scans allow us to visually
pinpoint the boundary of the microphase region. General
arguments indicate that fluctuations alter the nature of the
critical point, rendering the transition weakly first
order~\cite{brazovskii:1975}.

At this stage it is useful to directly observe representative
equilibrium configurations obtained during GEMC runs, as in
Figs.~\ref{fig:2dconfig}(a)-(c). Configurations below the
microphase transition show slightly irregular domains, due to
finite-temperature entropic fluctuations. For this system at
the densities and temperatures shown, compact cluster states
are more stable than extended (lamellar) states, which were not
observed. The second and third panels of
Fig.~\ref{fig:2dconfig}c illustrating configurations just above
and below the microphase line are particularly illuminating.
The dynamical behavior upon a rapid quench below the microphase
separation line is shown in the first two panels of
Fig.~\ref{fig:2dconfig}(d).\footnote{MD integration with step
$\Delta t$=$0.001$ was performed for systems sizes $N\geq 256$.
Cooling is done from a fluid configuration to the set
temperature by resampling velocities every $t_{cool}$=$10$. The
type of dynamics does not influence the qualitative results
strongly~\cite{foffi:2005b}, so other protocols are not
studied. Slower cooling rates allow the system to reach
configurations closer to equilibrium~\cite{sciortino:2004}. The
potential is truncated at $r_c$=$5$($3.3$) for $\xi$=$2$($1$),
which alters the microphase patterning~\cite{imperio:2004}, but
here the results do not depend on the form of the ground state,
since GEMC and MD runs use equal $r_c$.} The evolution of the
system from the second to the third panel of
Fig.~\ref{fig:2dconfig}(d) is extremely slow via conventional
MD. The dramatic increase in time scales upon cooling the
system below the microphase separation can be seen from the
self-intermediate scattering function $F_s(k,t)$ in the inset
of Fig.~\ref{fig:2dph}(b). Fluctuations on the order of the
characteristic domain length scale decay so slowly that the
structure is effectively dynamically arrested.

\begin{figure}
\center{\includegraphics[width=0.3\columnwidth]{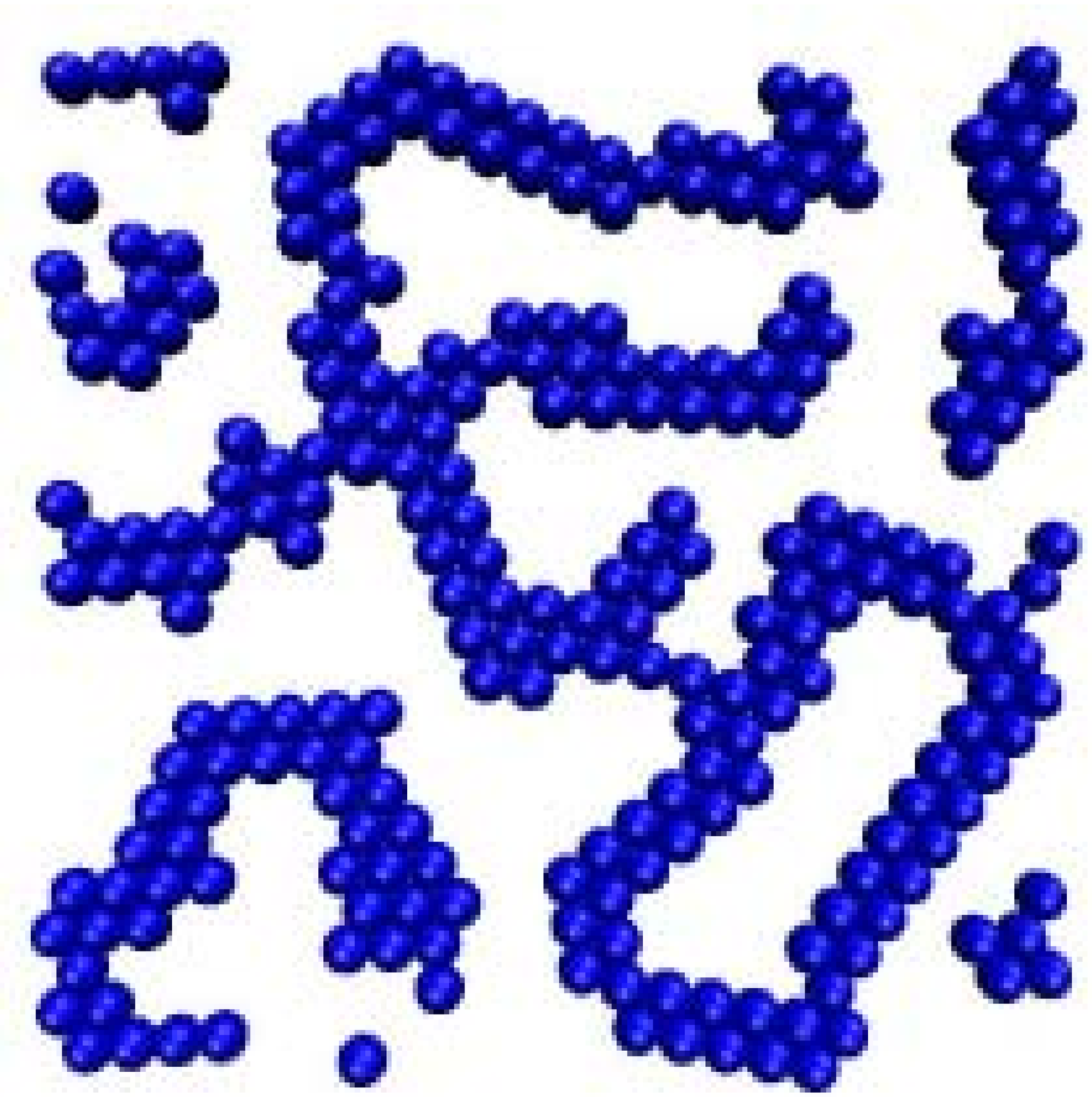}\includegraphics[width=0.3\columnwidth]{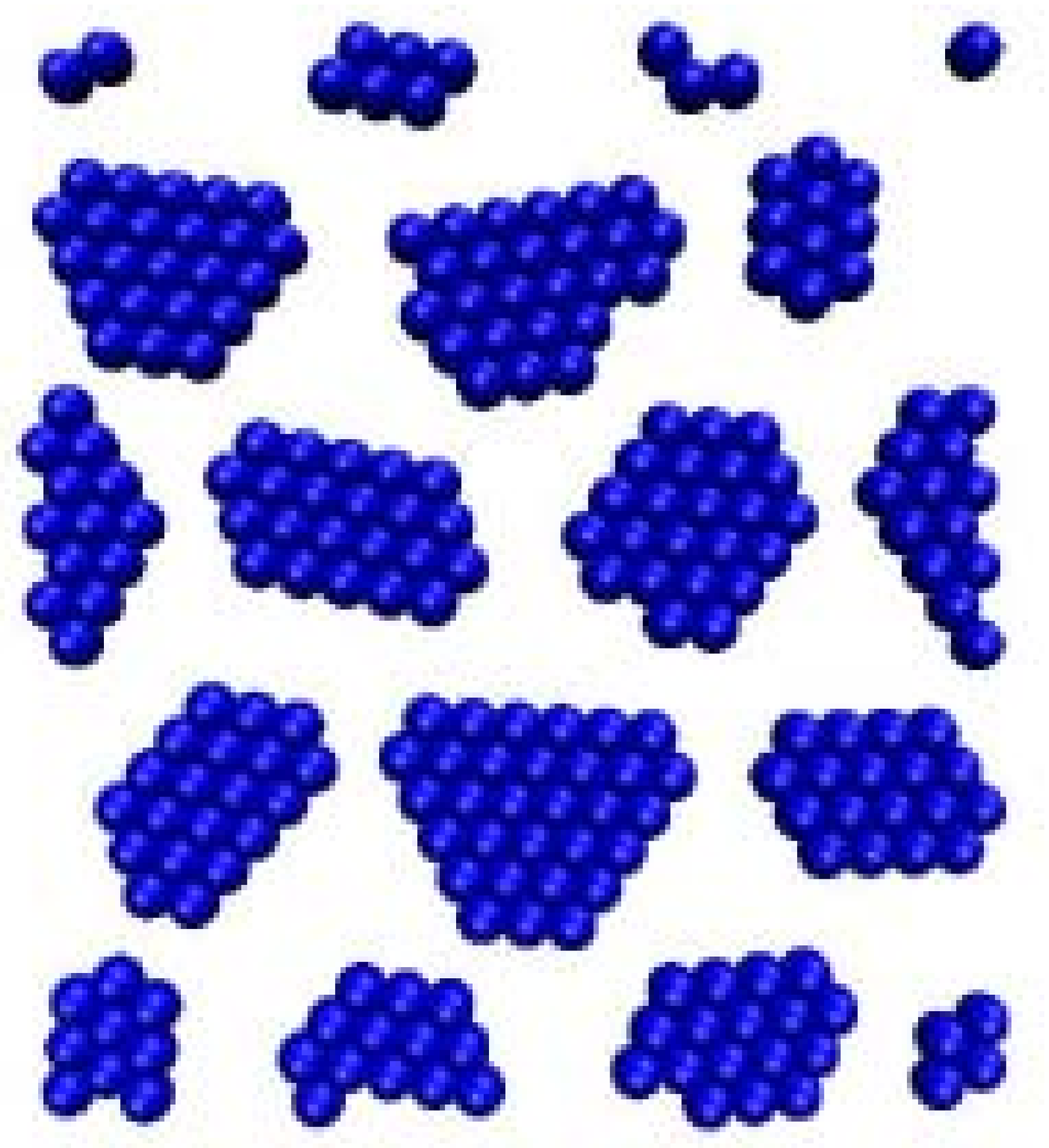}}\\
\caption[Two-dimensional GEMC ground state and gel
configurations of microphase-forming systems]{(Color online)
For $\rho$=$0.5$ ($\xi$=$2$, $A$=$0.2$) (a) a nearly arrested
gel configuration obtained by MD at $t$=$5\times 10^{5}$ after
a quench to $T$=$0.08$ and (b) an equilibrium configuration at
$T$=$0.08$ found by GEMC simulation.} \label{fig:2dgel}
\end{figure}
The dynamical behavior shown in Fig.~\ref{fig:2dconfig}(d) is
clearly relevant for the mechanism of gelation of the system.
In particular, consider slightly deeper quench at a lower
density than that shown in Fig.~\ref{fig:2dconfig}(d). In
Fig.~\ref{fig:2dgel}(a) a quench of the system is made at
$\rho$=$0.5$ to $T$=$0.08$. After rapid initial transient
dynamics, where a tenuous percolating structure is formed, the
dynamics becomes anomalously slow. This system may be
characterized as a gel.  Due to the depth of the quench, we did
not observe substantial structural coarsening [as is observed
in Fig.~\ref{fig:2dconfig}(d)] over the duration of the
simulations, although the energy of the system is slowly
evolving in time. The similar evolution is expected to be much
slower in the case of a polydisperse, three-dimensional system.
For comparison, the equilibrium state of the system (as found
via GEMC simulation) is also shown in Fig.~\ref{fig:2dgel}.
Note that the structure of the gel does not resemble the
equilibrium state in any regard, despite the fact that the
energy per particle in the gel is quite close to that of
particles in their cluster ground state. In addition, we find
via the GEMC method no evidence of extended microphase
structures at the density and temperature of
Fig.~\ref{fig:2dgel}. Regardless, the short-time critical-like
fluctuations that exist during the initial stages of microphase
separation are sufficient to generate configurations that span
space and evolve anomalously slowly, due to strong,
short-ranged bonding between particles.

The results presented above are clearly of importance for
understanding the general routes to gelation in colloidal
systems with competing short-ranged attraction and long-ranged
repulsion. We find, rather robustly, that gels may form via
arrested \emph{microphase} separation. This is analogous to the
purely attractive case, where gels are formed by arrest of
global phase separation. This picture is in contrast to that of
Refs.~\cite{sciortino:2004,wu:2004}, where gelation is a
consequence of near-equilibrium vitrification of clusters that
are stabilized by the repulsive interactions. We have
explicitly demonstrated here that, as in the purely attractive
case, the gel may be stabilized by the attraction between
particles. More interestingly, the gel structure bears no
similarity to the clusters that exist in thermodynamic
equilibrium.  We find that in the system studied here, the only
role of the repulsion is to select the symmetry of the ground
state(s). Our results are fully consistent with, but more
general than, the suggestion that disordered configurations of
anisotropic microphase structures are connected to
gelation~\cite{sciortino:2005,deCandia:2006,tarzia:2006}.
Indeed, we have directly demonstrated that extended microphase
structures are not involved during the formation of the gel in
Fig.~\ref{fig:2dgel}. Furthermore, preliminary studies in three
dimensions for the system studied in Ref.~\cite{sciortino:2004}
show that for rapid, deep quenches gels may be formed in the
same manner. This leads us to believe that both the routes
presented here and in Refs.~\cite{sciortino:2004,wu:2004} are
possible, and will likely depend intimately on the details of
the quench protocol. In general, lamellar and tubular
structures (as well as other microphase textures) may comprise
the building blocks of the disordered weak solids found in
colloidal systems with depletion attraction and long-ranged
charge repulsion, but our results demonstrate that they need
not.  The disordered lamellar and tubular examples are simply
some of the many ways in which microphase separation may be
arrested in a manner directly analogous to the arrest of global
phase separation. The collection of all such ways perhaps forms
the simplest generic route to the gel state at intermediate
volume fractions in attractive systems with charge frustration.
The search for more exotic routes which may be operative at
very low volume fractions and where charge repulsion plays a
more active role may be greatly facilitated by the use of the
GEMC approach as outlined here.

\vspace{0.15cm} This work was supported by grants No.
NSF-0134969 and No. FQRNT-91389 (P.C.). We thank E. Del Gado,
D. Frenkel, P. Lu, and K. Miyazaki for discussions and
comments.
\bibliography{LongRep}
\end{document}